\RequirePackage{fix-cm}
\documentclass[smallextended]{svjour3}       
\smartqed  
\usepackage{graphicx}

\usepackage[intlimits]{amsmath}
\usepackage{graphics}
\usepackage{amsfonts}
\usepackage{amssymb}
\usepackage{dsfont}
\usepackage{units}

\usepackage[usenames]{color}
\usepackage{colortbl}
\usepackage[colorlinks=true,linkcolor=blue,citecolor=blue,urlcolor=blue]{hyperref}

\usepackage{titlesec}
\usepackage{framed}
\usepackage{eurosym}
\usepackage{bbding}
\usepackage{cite}
\usepackage{slashed}

\usepackage{bm}
\usepackage{wasysym}
\usepackage{multirow}

        \usepackage[vcentermath]{youngtab}

\newcommand{\conjg}[1]{\ensuremath{\hspace{1pt}\overline{\hspace{-1pt}#1\hspace{-1pt}}}\hspace{1pt}}

\def\mA{\ensuremath{\mathcal{A}}}

\def\mD{\ensuremath{\mathcal{D}}}

\def\mK{\ensuremath{\mathcal{K}}}

\def\mS{\ensuremath{\mathcal{S}}}

\begin{document}

\title{Heavy baryon spectroscopy in a quark-diquark approach 
\thanks{This work is supported through the Portuguese Science Foundation FCT under project CERN/FIS-PAR/0023/2021, the FCT computing project 2021.09667.CPCA, and under the Doctoral Program Scholarship 2022.11964.BD.}
}



\author{André Torcato \and
        Ana Arriaga  \and \\
        Gernot Eichmann \and
        M. T. Peña
}


\institute{André Torcato \at
              LIP Lisboa, Av. Prof. Gama Pinto 2, 1649-003 Lisboa, Portugal \\
              Instituto Superior T\'ecnico, Universidade de Lisboa, 1049-001 Lisboa, Portugal\\
              \email{atorcato@lip.pt}           
           \and
           Ana Arriaga \at
              LIP Lisboa, Av. Prof. Gama Pinto 2, 1649-003 Lisboa, Portugal \\
              FCUL Lisboa, Campo Grande, 1749-016 Lisboa, Portugal \\
              \email{amarriaga@ciencias.ulisboa.pt}
           \and
           Gernot Eichmann \at
              Institute of Physics, University of Graz, NAWI Graz, Universitätsplatz 5, 8010 Graz, Austria \\
              \email{gernot.eichmann@uni-graz.at}
           \and
           M. T. Peña \at
              LIP Lisboa, Av. Prof. Gama Pinto 2, 1649-003 Lisboa, Portugal \\
              Departamento de Física and Departamento de Engenharia e Ciências Nucleares, Instituto Superior T\'ecnico, Universidade de Lisboa, 1049-001 Lisboa, Portugal \\
              \email{teresa.pena@tecnico.ulisboa.pt}
}

\date{Received: date / Accepted: date}

\maketitle

\begin{abstract}
We report progress on  calculations of the heavy-light baryons $\Sigma_c$ and $\Lambda_c$ and their excitations with $J^P = 1/2^+$ using functional methods. 
We employ a covariant quark-diquark approach, where the interaction amounts to a quark exchange between  quarks and diquarks 
and the ingredients are determined from the quark level. 
A partial-wave analysis reveals the presence of orbital angular momentum components in terms of \textit{p} waves, 
which are  non-relativistically suppressed.
\end{abstract}

\section{Introduction}
\label{intro}

Even though  experimental data on the heavy-baryon spectrum are still  sparse compared to light baryons,
the last decade has brought major progress in this area. 
Apart from intensely discussed tetraquark and pentaquark candidates,  by now almost 30 new heavy baryons 
have been found at the LHC~\cite{LHCb-FIGURE-2021-001-report,Chen:2022asf}, with the newest members added only  recently~\cite{LHCb:2023rtu}.
Many of these  states contain a single bottom quark ($\Sigma_b$, $\Lambda_b$, $\Xi_b$, $\Omega_b$), 
and singly-charmed ($\Lambda_c$, $\Xi_c$, $\Omega_c$) and doubly-charmed baryons ($\Xi_{cc}$)  have also been observed.
Clearly, this calls for a strong theory support in the study of heavy-baryon spectroscopy.
In addition to quark models (see e.g.~\cite{Liu:2019zoy,Brambilla:2019esw,Yang:2019lsg,Chen:2022asf} and references therein),
much theoretical progress in the field has been made using lattice QCD~\cite{Lewis:2008fu,Liu:2009jc,Briceno:2012wt,PACS-CS:2013vie,Alexandrou:2014sha,Brown:2014ena,Padmanath:2013zfa,Perez-Rubio:2015zqb,Alexandrou:2017xwd,Mathur:2018rwu,Bahtiyar:2020uuj}.

In this work we address heavy-baryon spectroscopy with functional methods, in particular  Dyson-Schwinger equations (DSEs),
Bethe-Salpeter equations (BSEs) and covariant Faddeev equations. 
While heavy-meson studies in this approach have well advanced~\cite{Rojas:2014aka,Fischer:2014cfa,Hilger:2014nma,Hilger:2017jti,Serna:2020txe,Yao:2021pdy}, which also includes investigations of four-quark states~\cite{Wallbott:2020jzh,Eichmann:2020oqt,Santowsky:2021bhy},
applications in the baryon sector have  so far mainly been developed for light and strange baryons, see~\cite{Eichmann:2016yit,Barabanov:2020jvn} for reviews.
Here the three-quark Faddeev equation and its reduction to a quark-diquark system 
 have proven  suitable to describe the light and strange baryon spectrum~\cite{Eichmann:2016hgl,Fischer:2017cte,Eichmann:2018adq,Liu:2022nku}.
Comprehensive studies of charm and bottom baryons using an equal spacing scheme have been performed with the three-quark Faddeev equation~\cite{Qin:2018dqp,Qin:2019hgk} and a contact-interaction approach~\cite{Yin:2021uom}, 
but so far direct calculations of singly- or doubly-charmed baryons are not yet available.
To establish a common description of light and heavy baryons using the same microscopic ingredients,
it is thus desirable to extend the existing tools  to the heavy-baryon sector.

Here we  report on progress using the quark-diquark approach, which we apply to investigate 
the singly-charmed baryons $\Sigma_c$ and $\Lambda_c$ with $J^P = 1/2^+$.
We describe the approach in Sec.~\ref{sec:qdq}, present first results in Sec.~\ref{sec:results} and
briefly summarize in Sec.~\ref{sec:summary}.

\section{Quark-diquark BSE} \label{sec:qdq}

\begin{figure}
\centering
  \includegraphics[width=0.75\textwidth]{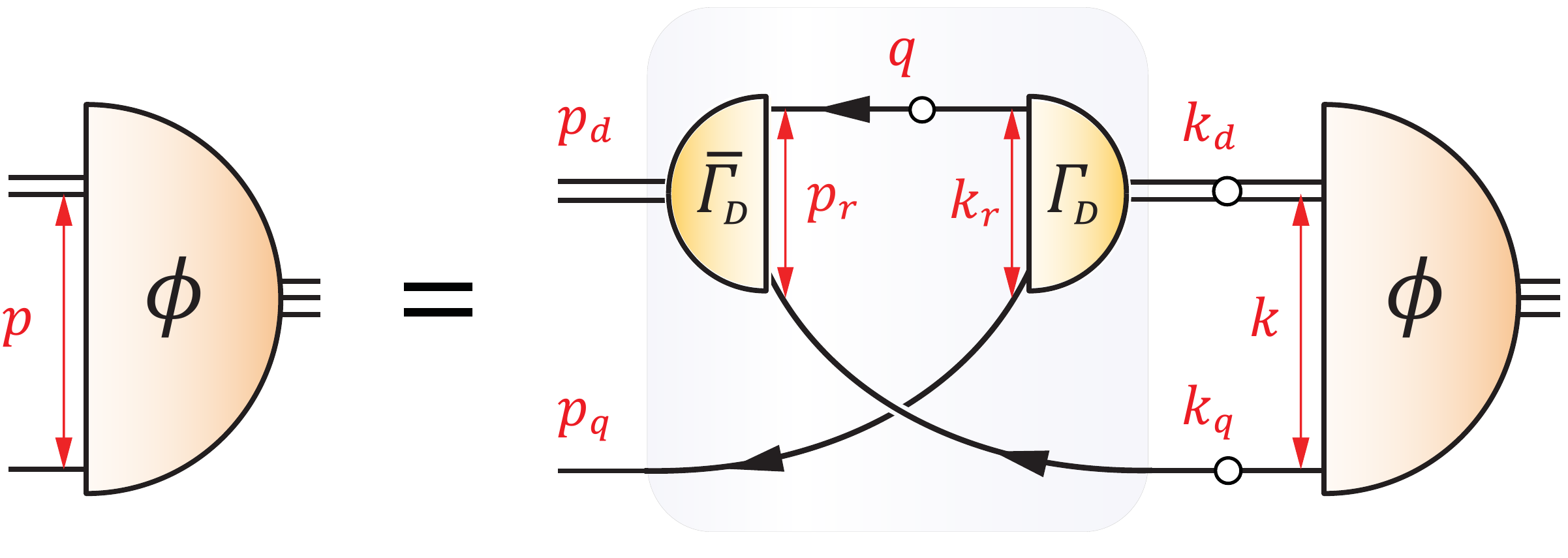}
\caption{Quark-diquark BSE from Eq.~\eqref{qdq}}
\label{fig:qdq}
\end{figure}

The reduction of the covariant three-body Faddeev equation to a quark-diquark BSE is described in detail in~\cite{Eichmann:2016yit}.
Upon neglecting irreducible three-quark contributions, one employs a diquark ansatz for the two-quark scattering matrix
through a sum over diquark correlations. The sum is dominated by the diquarks with smallest masses, which
are the positive-parity scalar and axialvector diquarks followed by the negative-parity pseudoscalar and vector diquarks.
The resulting quark-diquark BSE is shown in Fig.~\ref{fig:qdq} and reads
\begin{equation}\label{qdq}
    \phi^{(\mu)}_{\alpha\sigma}(p,P) = \int \!\! \frac{d^4k}{(2\pi)^4} \, \mathbf{K}^{(\mu\nu)}_{\alpha\beta} \, S_{\beta\gamma}(k_q) \, D^{(\nu\rho)}(k_d) \, \phi^{(\rho)}_{\gamma\sigma}(k,P)\,,
\end{equation}
where $\mathbf{K}$ is the quark-diquark interaction kernel given by
\begin{equation}\label{qdq-kernel}
    \mathbf{K}^{(\mu\nu)} = \Gamma_\text{D}^{(\nu)}(k_r,k_d)\,S^T(q)\,\conjg\Gamma_\text{D}^{(\mu)}(p_r,p_d)\,.
\end{equation}

\newpage

Here, $\phi(p,P)$ is the quark-diquark Bethe-Salpeter amplitude depending on the relative momentum $p$ and total momentum $P$ with $P^2 = -M^2$, where $M$ is the mass of the baryon.
The remaining momenta can be inferred from Fig.~\ref{fig:qdq}. $S$ is the dressed quark propagator, $D$ the diquark propagator and $\Gamma_\text{D}$ the diquark amplitude, with $\conjg\Gamma_\text{D}$ its charge conjugate.
The Greek subscripts are Dirac indices and the Lorentz indices appear only for (axial-)vector diquarks.

In Ref.~\cite{Eichmann:2016hgl} this approach was employed to calculate the light baryon excitation spectrum.
The ingredients of the equation -- the quark propagator, diquark propagator and diquark amplitudes -- were computed
from their  DSEs and BSEs in a rainbow-ladder truncation, which allows for a direct
comparison with the original three-body Faddeev equation using the same input. It turns out that the three-body
and quark-diquark calculations yield very similar spectra, which supports the idea of diquark clustering inside baryons~\cite{Barabanov:2020jvn}. 
Moreover, while the  $J^P = 1/2^+$ nucleon and  $J^P=3/2^+$ $\Delta$ channels
agree well with experiment, the remaining $J^P$ channels are too strongly bound in both approaches. This is due to
the higher-lying pseudoscalar and vector diquarks, whose scalar and axialvector meson partners are too strongly bound in rainbow-ladder.
However, one can simulate beyond rainbow-ladder effects by introducing one parameter that pushes the axialvector mesons into the experimental ballpark;
the corresponding diquarks then follow course and the resulting baryon spectrum turns out to be in 1:1 agreement with experiment~\cite{Eichmann:2016hgl}.
Preliminary results for the strange baryon spectrum using the same strategy can be found in Refs.~\cite{Fischer:2017cte,Eichmann:2018adq}.

\begin{table}[!b]
\caption{$S_3$ classification of light and singly-charmed baryons}
\label{tab:flavor}       
\begin{tabular}{l @{\qquad\quad} llll @{\qquad\quad} lll}
\hline\noalign{\smallskip}
        & $uuu$         & $uud$      & $udd$      & $ddd$      & $uuc$           & $udc$         & $ddc$  \\ \noalign{\smallskip}\hline\noalign{\smallskip}
$\mS$   & $\Delta^{++}$ & $\Delta^+$ & $\Delta^0$ & $\Delta^-$ & $\Sigma_c^{++}$ & $\Sigma_c^+$  & $\Sigma_c^0$ \\[0.5mm]
$\mD_1$ &               & $p$        & $n$        &            & $\Sigma_c^{++}$ & $\Sigma_c^+$  & $\Sigma_c^0$ \\[0.5mm]
$\mD_2$ &               &            &            &            &                 & $\Lambda_c^+$ & \\[0.5mm]
$\mA$   &               &            &            &            &                 & $\Lambda_c^+$ & \\
\noalign{\smallskip}\hline
\end{tabular}
\end{table}

The goal of the present work is to extend this approach to charmed baryons. 
We focus in particular on the $\Sigma_c$ and $\Lambda_c$ baryons with quark content $nnc$, where $n=u,d$ stands for light valence quarks and $c$ for charm.
As shown in~Table~\ref{tab:flavor}, their flavor wave functions can be arranged in symmetric singlets $\mS$, mixed-symmetric doublets $\mD$ and antisymmetric singlets $\mA$ of the permutation group $S_3$ 
(see e.g. Ref.~\cite{Eichmann:2022zxn} for the explicit construction).
For example, in the case of the $\Sigma_c^{++}$ with quark content $uuc$ one obtains a singlet and a doublet,
\begin{equation} \label{sd}
  \mS = \frac{1}{\sqrt{3}}\left( uuc + \{uc\}u\right), \qquad
  \mD_1 = \left( \frac{1}{\sqrt{2}} [uc]u \atop \frac{1}{\sqrt{6}} \left( 2uuc - \{uc\}u \right) \right),
\end{equation}
where we abbreviated $[uc]=uc-cu$ and $\{uc\} = uc+cu$.
In the $SU(4)_F$ arrangement $\mathbf{4}\otimes\mathbf{4}\otimes\mathbf{4} = \mathbf{20}_S \oplus \mathbf{20}_{M_A} \oplus \mathbf{20}_{M_S} \oplus \mathbf{4}_A$ 
the singlet belongs to $\mathbf{20}_S$ (which contains the decuplet baryons in the three-flavor case) 
and the doublet components to $\mathbf{20}_{M_A}$ and  $\mathbf{20}_{M_S}$ (containing the octet baryons).
The full  wave function for a given baryon is  then the combination of flavor, color and Dirac-momentum components, but
because  $SU(4)_F$  is broken the Dirac-momentum parts do not show a particular symmetry and 
the  components with different flavor wave functions ($\mS$ and $\mD_1$ for the $\Sigma_c$) can mix.

\begin{figure}
\centering
  \includegraphics[width=0.9\textwidth]{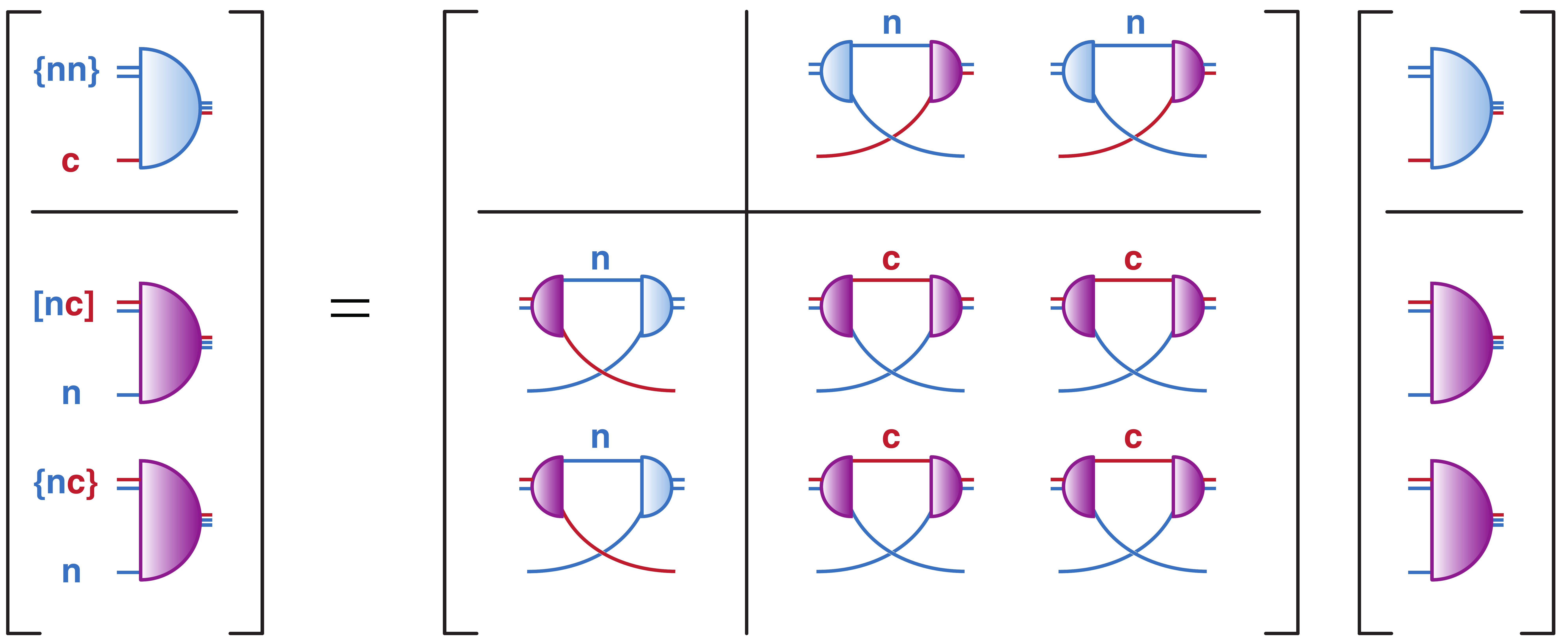}
\caption{Coupled quark-diquark BSEs for the $\Sigma_c$}
\label{fig:qdq-coupled}
\end{figure}

In the quark-diquark approach Eq.~\eqref{qdq} becomes a coupled system of equations in flavor space, which is illustrated
in Fig.~\ref{fig:qdq-coupled} for the $\Sigma_c$ baryons.
For quark content $nnc$ there are eight possible diquark flavor wave functions,
which are (anti-)symmetric under quark exchange:
\begin{equation}
   [ud], \quad [uc], \quad [dc], \qquad\qquad uu, \quad \{ud\}, \quad dd, \quad \{uc\}, \quad \{dc\}\,.  
\end{equation}
With $\mathbf{4}\otimes\mathbf{4} = \mathbf{6}_A \oplus \mathbf{10}_S$ the first three belong to $\mathbf{6}_A$ and
the remaining five to  $\mathbf{10}_S$.
In the following we restrict ourselves to scalar and axialvector diquarks.
For identical quark masses, the antisymmetric $[nn]$ configuration corresponds to scalar diquarks
and the symmetric ones $\{nn\}$, $\{cc\}$ to axialvector diquarks. For heavy-light diquarks 
this is no longer necessarily the case but we still identify $[nc]$ with the scalar diquark and $\{nc\}$ with the axialvector one.
The $\Sigma_c$ then only features the components $\{nn\}c$, $[nc]n$ and $\{nc\}n$ but not $[nn]c$
due to its isospin,
which leads to the structure in Fig.~\ref{fig:qdq-coupled}.
For the $\Lambda_c$, on the other hand, $[nn]c$ appears instead of $\{nn\}c$.
For a given baryon, it is then straightforward to determine the flavor factors for the various kernel components in Fig.~\ref{fig:qdq-coupled}
by taking the flavor traces in the quark-diquark BSE.  

The $SU(4)_F$ breaking implies that the singlet and doublet components in Table~\ref{tab:flavor} can mix, which is already implemented in the equation:
The kernel is identical for both components except that for the singlet component of the $\Sigma_c$  the rows and columns for $[nc]n$ are absent.
Therefore, once the equation is solved using the full kernel,  the singlet- and doublet-like components 
dynamically appear as ground or excited states for a baryon with given $J^P$, which is analogous to the situation for strange baryons~\cite{Eichmann:2018adq}.

We employ the full Dirac-momentum structure of the quark-diquark amplitudes allowed by Poincaré covariance~\cite{Oettel:1998bk,Eichmann:2016yit}.
For $J^P=1/2^+$ baryons this comprises eight tensors, two for scalar
and six for axialvector diquarks,
see e.g. Eqs.~(11-12) and Tables I--II in Ref.~\cite{Eichmann:2011aa} for details:
\begin{equation}\label{dirac}
\begin{split}
   \phi(p,P) &= \sum_{k=1}^2 f_k(p^2,z)\,\tau_k(p,P)\,\Lambda_+(P)\,, \\
   \phi^\mu(p,P) &= \sum_{k=3}^8 f_k(p^2,z)\,\tau_k^\mu(p,P)\,\gamma_5 \Lambda_+(P)\,.
\end{split}
\end{equation}
Here, $\Lambda_+(P) = (\mathds{1} + \hat{\slashed{P}})/2$ is the positive-energy projector, a hat denotes a normalized four-momentum,
and $z=\hat{p}\cdot\hat{P}$ is the cosine of the four-dimensional angle.
The tensors $\tau_k(p,P)$ can be arranged as eigenstates of the quark-diquark orbital angular momentum in the baryon's rest frame,
which leads to $s$-wave  ($l=0$), $p$-wave  ($l=1$) and $d$-wave components ($l=2$);
for $J=3/2$ baryons also $f$ waves contribute.
The dressing functions $f_k(p^2,z)$ are the dynamical output of the equation.
After taking Dirac, color and flavor traces and performing a Chebyshev decomposition in $z$, 
the quark-diquark BSE turns into an eigenvalue equation
\begin{equation}
    \mK(P^2)\,\psi_i(P^2) = \lambda_i(P^2)\,\psi_i(P^2)\,,
\end{equation}
where the masses of the ground and excited states are determined from the condition $\lambda_i(P^2=-M_i^2) = 1$.
The dimension $(6+2+6) \times N_p \times N_\text{cheb}$ of the kernel matrix in Fig.~\ref{fig:qdq-coupled} is then typically of the order of a few thousand depending on the precision, where $N_p$
is the number of grid points in $p^2$ and $N_\text{cheb}$ the number of Chebyshev moments.

Assuming isospin symmetry, the ingredients of the quark-diquark BSE are the $n$ and $c$ quark propagators
and the $[nn]$, $\{nn\}$, $[nc]$ and $\{nc\}$ diquark amplitudes and propagators, which must be determined beforehand.
We employ the same strategy as for light and strange baryons and calculate them from their DSEs and BSEs
in a rainbow-ladder truncation using the Maris-Tandy interaction~\cite{Maris:1999nt}; see Refs.~\cite{Eichmann:2009zx,Eichmann:2016hgl} for details.
The implementation of charm quarks, however, leads to several challenges:

    \smallskip
    {\tiny$\blacksquare$} 
    The quark propagators appearing in the diquark and quark-diquark BSEs are sampled on parabolas in the complex momentum plane.
    In particular, the charm-quark propagator is probed deep in the timelike region.
    To obtain a numerically reliable DSE solution in the complex plane we employ the
    Cauchy integration method developed in~\cite{Fischer:2005en,Krassnigg:2009gd}.

    \smallskip
    {\tiny$\blacksquare$} 
    The BSE eigenvalue spectrum for the $[nc]$ and $\{nc\}$ diquarks, which determines the diquark masses,
    is  contaminated by complex conjugate eigenvalues. This is a typical problem for heavy-light systems,
    and to resolve it  we resort to the spectral reconstruction method described in Ref.~\cite{Eichmann:2016nsu}: we reconstruct the propagator matrix
    from its positive eigenvalues only, which through a Cholesky decomposition  guarantees
    that the final BSE eigenvalues are real and the eigenvectors (the diquark amplitudes)  orthogonal.
    We employ the same strategy also for the quark-diquark BSE which faces similar issues.

\begin{figure}
\centering
  \includegraphics[width=0.95\textwidth]{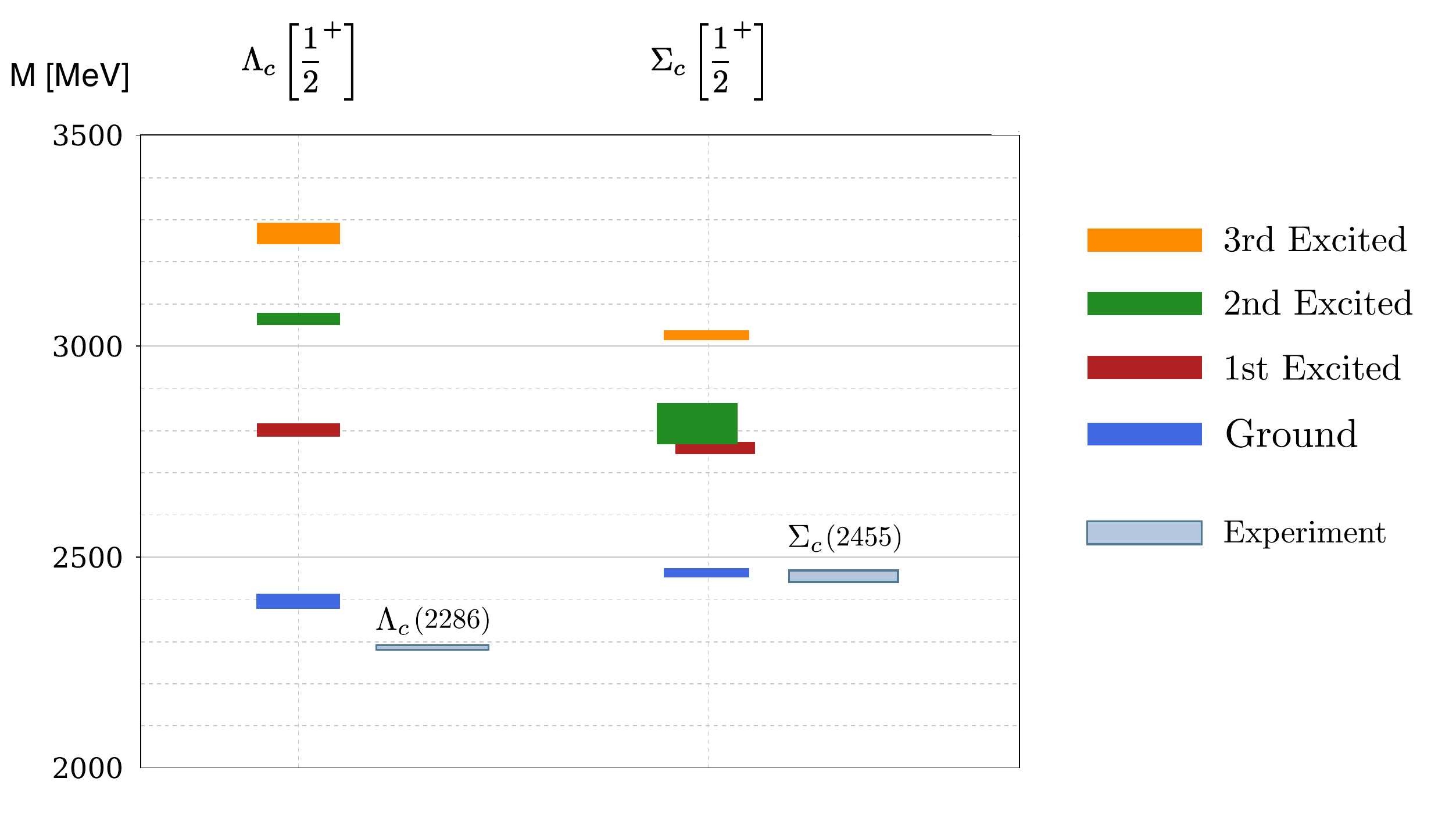}
\caption{Results for the singly-charmed baryon spectrum including $\Lambda_c$ and $\Sigma_c$ states compared to experiment~\cite{Workman:2022ynf}}
\label{fig:spectrum}
\end{figure}    
    
    \smallskip
    {\tiny$\blacksquare$}
    The masses of the heavy-light diquarks $[nc]$ and $\{nc\}$ turn out to lie just above the quark contour limit, where the diquark BSE probes the quark propagators beyond their leading singularities in the timelike complex plane.
    In the absence of residue calculations, the diquark masses can be extrapolated to their onshell values
    by altering the coupling in the diquark BSEs. However, this still complicates 
    the calculation of the diquark amplitudes (and
    in particular their canonical normalizations) which enter in the quark-diquark BSE. 
    For this reason we compensate the change by readjusting the diquark normalization
    and fitting it to the $\Sigma_c$ ground state.

    \smallskip
    {\tiny$\blacksquare$}
    In the quark-diquark BSE the quark and diquark propagators are also sampled in the complex plane, which leads
    to analogous singularity restrictions. This becomes particularly severe for the heavy-light baryon spectrum, where all masses
    above $\sim 2.4$ GeV result from extrapolations, in particular also the ground states.
    Also in this case the problem can in principle be solved  by residue calculations, but this would pose a substantial numerical effort
    which we do not attempt here. On the other hand, there are examples demonstrating that (in the absence of physical thresholds)
    extrapolations agree with direct calculations where such a comparison is possible, see e.g.~\cite{Huber:2021yfy}.

\begin{figure}
\centering
  \includegraphics[width=0.7\textwidth]{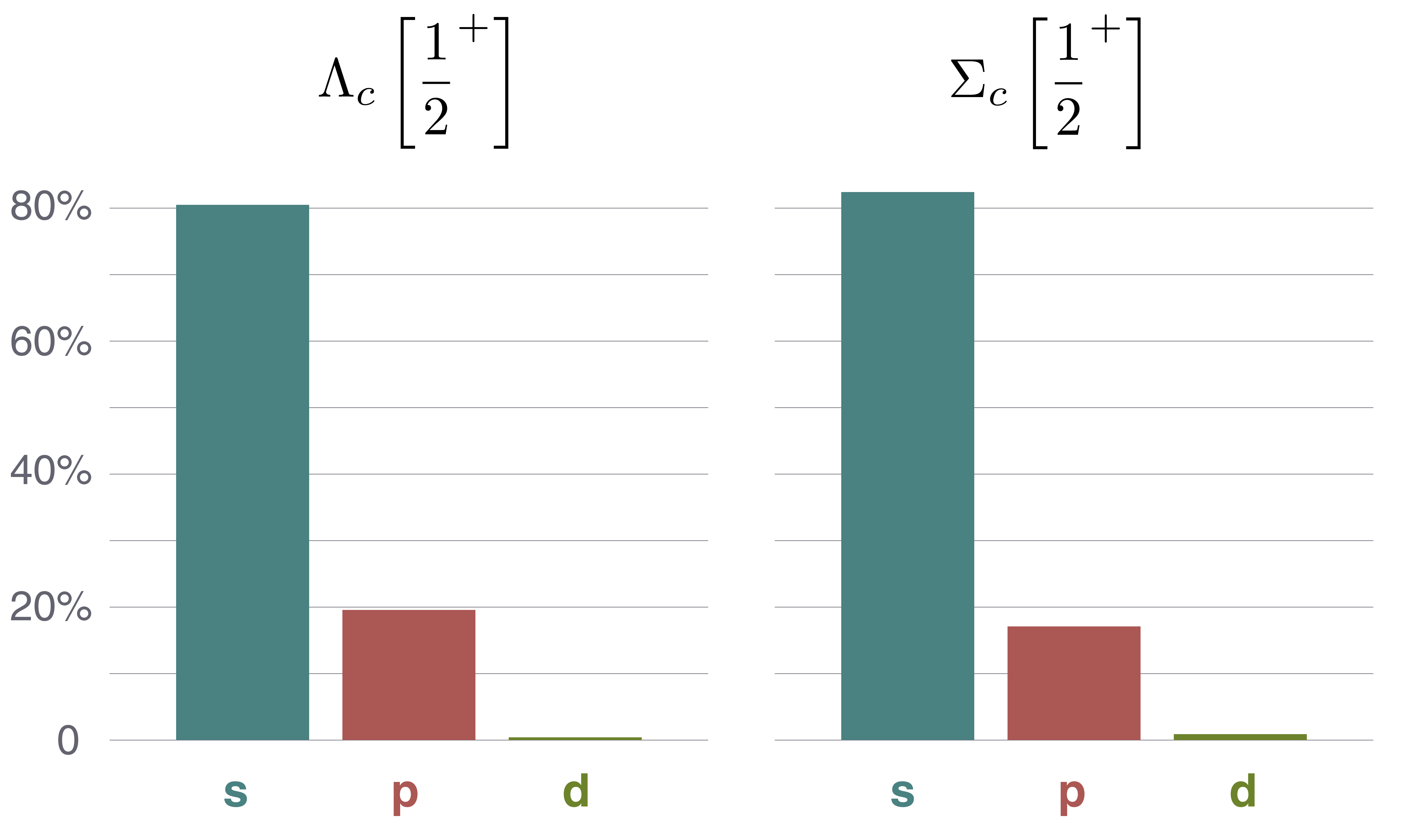}
\caption{Partial-wave contributions to the $\Lambda_c$ and $\Sigma_c$ states, given as percentages contributing to the total baryon normalization}
\label{fig:spdf}
\end{figure}     

\begin{figure}[!b]
\centering
  \includegraphics[width=0.7\textwidth]{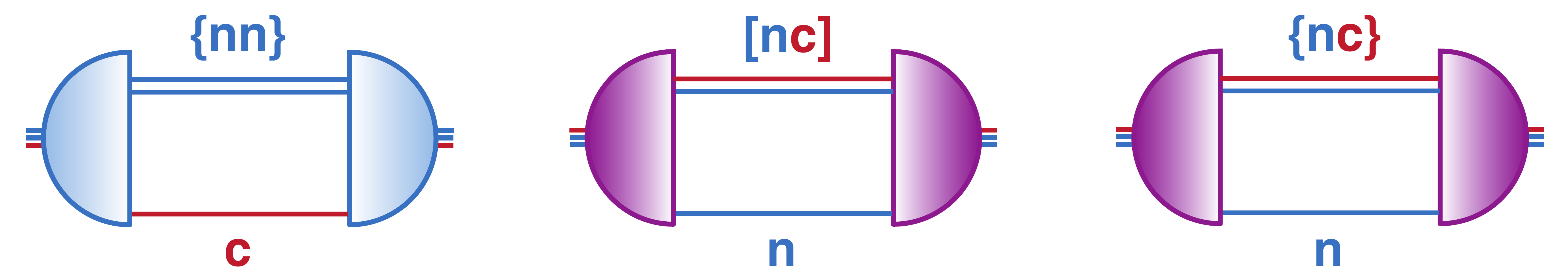}
\caption{Overlap integrals that enter in the baryon's normalization, exemplified for the $\Sigma_c$}
\label{fig:normalization}
\end{figure}  
    
\section{Results} \label{sec:results}

  We show results for the $\Lambda_c (1/2^+)$ and $\Sigma_c (1/2^+)$ mass spectra in Fig.~\ref{fig:spectrum}.
  As mentioned earlier,  all masses are extrapolated and the bars correspond to the extrapolation errors.
  The $\Sigma_c$ states mirror the  results in Refs.~\cite{Fischer:2017cte,Eichmann:2018adq} for the $\Sigma$ baryons,
  which resemble a superposition of octet and decuplet spectra.
  The $1/2^+$ channel features a nucleon-like ground state corresponding to the $\Sigma_c(2455)$ and a Roper-like first excitation, 
  and the $\Lambda_c$ channel resembles a superposition of octet and singlet spectra,
  with the $\Lambda_c(2286)$ as the nucleon-like ground state.
  Some of the higher-lying excitations also appear to extrapolate into the same mass region, 
  but these are numerically less reliable and due to the large extrapolation errors we do not include them in the plot.
  
  Fig.~\ref{fig:spdf} shows the composition of the ground states in terms of orbital angular momentum components,
  in particular the contributions from the various tensors in Eq.~\eqref{dirac} to the baryons' canonical normalization.
  The latter is proportional to the overlap integrals shown in Fig.~\ref{fig:normalization}, 
  which depend on the dressed propagators and calculated amplitudes.
  Both $\Lambda_c$ and $\Sigma_c$ feature close to $20\%$ $p$ waves, which are relativistic contributions.
  This can be compared to the nucleon ground state, where the $p$-wave admixture is $\sim 33\%$~\cite{Eichmann:2011vu,Eichmann:2016hgl}.
  Similar observations also hold for other heavy baryons obtained with the three-quark Faddeev equation~\cite{Qin:2018dqp},
  which implies that relativity still plays  an important role for baryons made of charm quarks.

  \pagebreak
  
\begin{figure}
\centering
  \includegraphics[width=1\textwidth]{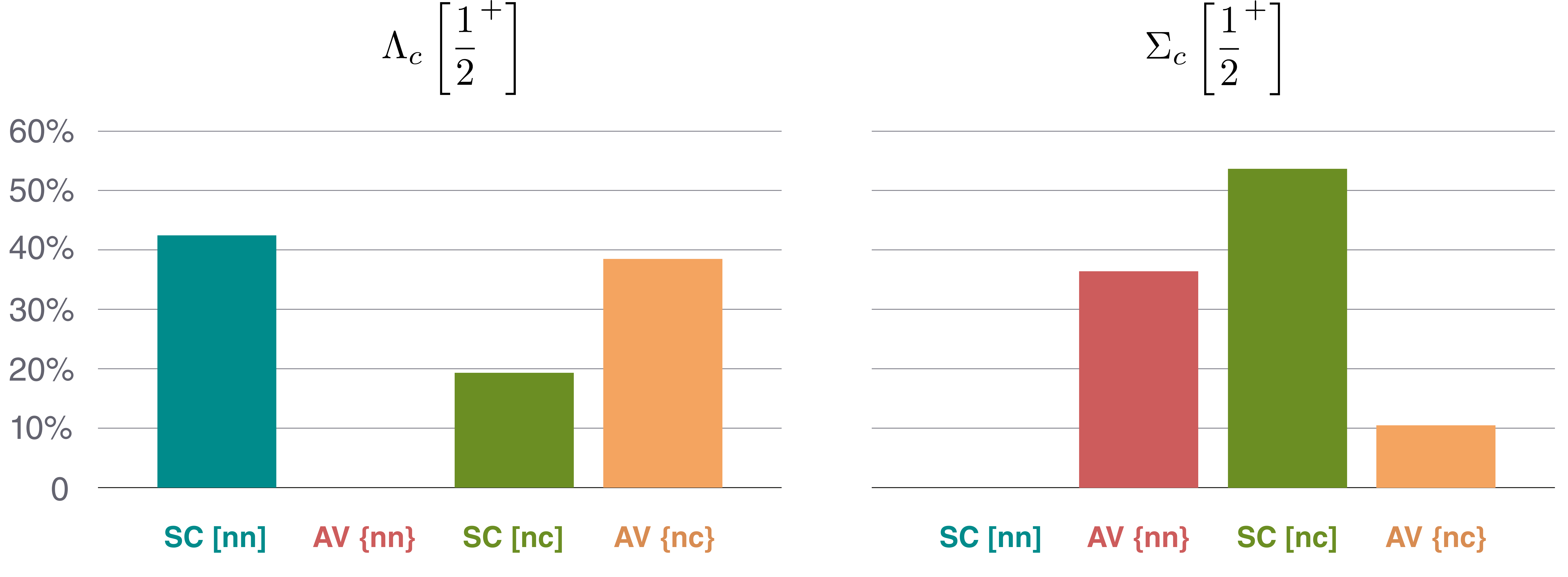}
\caption{Contributions to the $\Lambda_c$ and $\Sigma_c$ normalizations from the different diquark channels}
\label{fig:dq-contributions}
\end{figure}

  Finally, in Fig.~\ref{fig:dq-contributions} we show the composition of the $\Lambda_c$ and $\Sigma_c$ in terms of their diquark contributions,
  which are again obtained from the integrals in Fig.~\ref{fig:normalization}.
  One can  see that the two states have a very different internal structure: 
  The $\Lambda_c$ has similarly strong scalar-diquark $[nn]c$
  and axialvector $\{nc\}n$ components, followed by the scalar $[nc]n$ component, whereas the axialvector-diquark $\{nn\}c$ contribution is absent.
  The $\Sigma_c$, on the other hand, is dominated by the scalar $[nc]n$ component and has smaller axialvector components, whereas 
  the scalar $[nn]c$ component is absent.  
  It is clear from these findings
  that a scalar diquark component alone would not be sufficient to describe these baryons.
  
\section{Summary and Outlook} \label{sec:summary}  
  
  In this work we described the extension of the quark-diquark approach of Ref.~\cite{Eichmann:2016hgl} to heavy baryons,
  in particular to the $\Sigma_c$ and $\Lambda_c$ with $J^P = 1/2^+$.
  Calculations for heavy baryons with different quantum numbers and different quark content ($nsc$, $ssc$, $ncc$, $scc$, $ccc$)
   are already underway. In the longer term, further interesting extensions 
  are the generalization to the three-quark Faddeev equation and the systematic implementation of kernels beyond rainbow-ladder,
  where the underlying correlation functions are calculated fully self-consistently. Such studies will be the subject of future work.

\begin{acknowledgements}
We are grateful to Christian Fischer for discussions.
\end{acknowledgements}

\bibliographystyle{spmpsci_unsrt}

\bibliography{baryons}

\end{document}